\documentclass[journal]{IEEEtran}
\usepackage{graphicx,here,subfigure}

\begin{document}
\title{Construction of the first full-size GEM-based prototype for the CMS high-$\eta$ muon system}
%
%

\author{D.~Abbaneo,
        S.~Bally,
        H.~Postema,
	A.~Conde~Garcia,
	J.~P.~Chatelain,
	G.~Faber,
	L.~Ropelewski,
	S.~Duarte~Pinto,
	G.~Croci,
	M.~Alfonsi,
	M.~Van~Stenis,
	A.~Sharma,~\IEEEmembership{Senior Member,~IEEE},
	L.~Benussi,
	S.~Bianco,
	S.~Colafranceschi*,
	F.~Fabbri, 
	L.~Passamonti,
	D.~Piccolo,
	D.~Pierluigi,
        G.~Raffone,
	A.~Russo,	
	G.~Saviano,
	A.~Marinov,
	M.~Tytgat,~\IEEEmembership{Member,~IEEE},
	N.~Zaganidis,
	M.~Hohlmann,~\IEEEmembership{Member,~IEEE},
	K.~Gnanvo,
        M.G.~Bagliesi,
        R.~Cecchi,
	N.~Turini,
	E.~Oliveri,
	G.~Magazz\`u,
	Y.~Ban,
	H.~Teng,
	J.~Cai.
\\
RD51-NOTE-2010-008
\\
{\scriptsize This work has been submitted to the IEEE Nucl. Sci. Symp. 2010 for
publication in the conference record. Copyright may be transferred
without notice, after which this version may no longer be available. }
\thanks{Manuscript received November 18, 2010}
\thanks{D.~Abbaneo, S.~Bally, H.~Postema, A.~Conde~Garcia, J.-P.~Chatelain, G.~Faber, L.~Ropelewski, S.~Duarte~Pinto, G.~Croci, M.~Alfonsi, M.~Van~Stenis, A.~Sharma,
       are with Physics Department, CERN, Geneva, Switzerland}
\thanks{S.~Colafranceschi is with CERN, Geneva, Switzerland and Laboratori Nazionali di Frascati dell'INFN, Frascati, Italy and Sapienza Universit\`a di Roma - Facolta' Ingegneria}
\thanks{L.~Benussi, S.~Bianco, F.~Fabbri, L.~Passamonti, D.~Piccolo, D.~Pierluigi, G.~Raffone are with Laboratori Nazionali di Frascati dell'INFN, Frascati, Italy}
\thanks{G.~Saviano is with Sapienza Universit\`a di Roma - Facolta' Ingegneria, Rome, Italy and Laboratori Nazionali di Frascati dell'INFN, Frascati, Italy}
\thanks{A.~Marinov, M.~Tytgat, N.~Zaganidis are with the Department of Physics and Astronomy, Universiteit Gent, Gent, Belgium}
\thanks{M.~Hohlmann, K.~Gnanvo are with Dept. of Physics and Space Sciences, Florida Institute of Technology, Melbourne, FL, United States of America}
\thanks{M.G.~Bagliesi, R.~Cecchi, N.~Turini, E.~Oliveri, G.~Magazz\`u are with INFN, Sezione di Pisa, Universit\`a Degli Studi di Siena, Siena, Italy}
\thanks{Y.~Ban, H.~Teng, J.~Cai are with Peking University, Beijing, China}
\thanks{* Corresponding author: stefano.colafranceschi@cern.ch}
}

\maketitle
\pagestyle{empty}
\thispagestyle{empty}

\begin{abstract}
In view of a possible extension of the forward CMS muon detector system and future LHC luminosity upgrades, 
Micro-Pattern Gas Detectors (MPGDs) are an appealing technology. 
They can simultaneously provide precision tracking and fast trigger information, as well as sufficiently fine segmentation to cope with high particle rates
 in the high-eta region at LHC and its future upgrades. 
We report on the design and construction of a full-size prototype for the CMS endcap system, the largest Triple-GEM detector built to-date. 
We present details on the 3D modeling of the detector geometry, 
the implementation of the readout strips and electronics, and the detector assembly procedure.
\end{abstract}


\section{Introduction}
\IEEEPARstart{T}{}he CMS\cite{:2008zzk} muon system relies on three detector technologies: Drift Tubes (DT), Cathode Strip Chambers (CSC) and Resistive Plate Chambers\cite{Santonico:1994dk} (RPC).
The DT and CSC provide precision tracking functions, and RPCs provide fast trigger thanks to their excellent time resolutions.
Up to $\eta
\footnote{
$\eta$ is the pseudorapidity defined by:
  \begin{equation}
    \eta  = \frac{1}{2} \ln\left({ \frac {|\vec{p}|+p_L}{|\vec{p}|-p_L} }\right)
  \end{equation}
where $p_L$ is the component of the momentum $p$ along the beam axis. 
}
=1.6$ the forward muon region is instrumented with RPCs.

For region with high $|\eta| > 1.6$, during the CMS commissioning and construction, several concerns were raised on whether
 RPCs would be able to sustain the very hostile environment of the high-$\eta$ region; it was decided not to instrument this area at all.
Gas Electron Multipliers (GEMs)\cite{Sauli:1997qp} are an interesting technology for the future upgrade of the forward region of the muon system since 
they can provide precision tracking and fast trigger information simultaneously: moreover they can be designed with sufficiently fine segmentation to 
cope with high particle rates at LHC and its upgrades in the $\eta$ region to be covered.

\section{Candidate Technology}
We propose GEM technology as a candidate for the upgrade of the CMS muon system. Since 2009 we have been working on a feasibility study 
on the use of micro-pattern gas detectors (MPGD).
A full-size working chamber with dimensions $990\rm{mm} \times (220-455)\rm{mm}$, has been successfully built and preliminarily tested.

  \begin{figure}[H]
  \centering
  \includegraphics[width=3.5in]{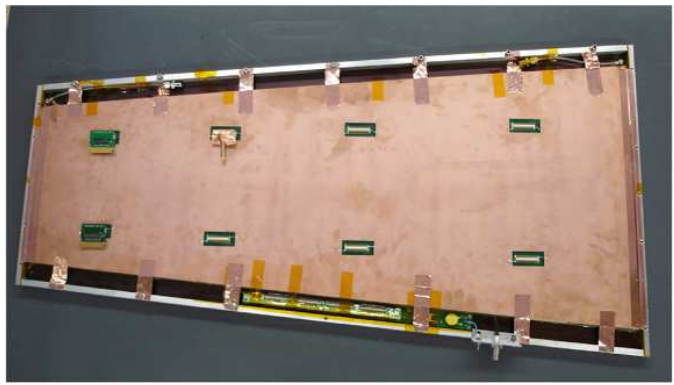}
  \caption{GE1/1 prototype chamber, dimensions: $990\rm{mm} \times (220-455)\rm{mm}$.}
  \label{ge11}
  \end{figure}
We started designing the chamber prototype at the end of 2009 following a very tight schedule that allowed to complete the construction of the first detector in October 2010.
As a starting point we developed a 3D detailed CATIA\cite{catia} model of a full-scale detector, to support the construction of a mock-up and functional prototype. 
This design was optimized with several studies focused on the mechanical integration (routing of cables, gas system) and detector performance such as gas flow simulations, readout and electronics.

\section{CMS high $\eta$ environment}
The high-$\eta$ region ($|\eta| > 1.6$) presents hostile conditions, with a particle fluence of several hundreds $\rm{Hz/cm^{2}}$ for an LHC luminosity of $10^{34} \rm{cm^{2}s^{-1}}$. 
Moreover, the particle rate might increase up to several $\rm{kHz/cm^2}$ depending on the LHC upgrade scenarios. 
In addition to this we must also consider the rates of thermal neutrons, low energy 
photons, and $\gamma$'s produced in a hadron collider such as the LHC. 
Hence there are several stringent requirements for a detector to be installed eventually in the high-$\eta$ region. 
This is the reason why the forward muon trigger system is equipped with RPC detectors only at $|\eta|< 1.6$. 
The high-$\eta$ region is presently vacant and presents an opportunity to be instrumented with a detector technology 
that could sustain the hard environment and be suitable for operations at the LHC and its future upgrades.

\section{Detector Construction}
\begin{figure}[H]
  \centering
  \includegraphics[width=3in]{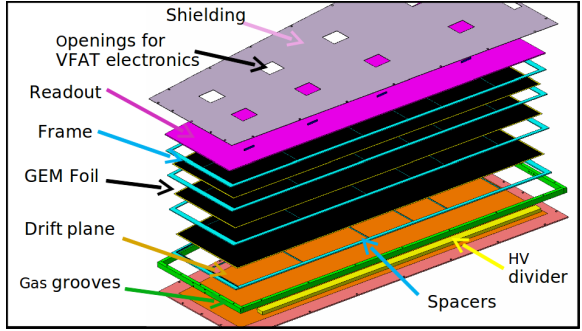}
  \caption{GE1/1 exploded CAD view.}
  \label{ge11cad}
  \end{figure}

The exploded CAD view, in Fig.\ref{ge11cad}, shows the mechanics derived from the originally planned RPC detector RE1/1. 
The proposed "GEM Endcap Station 1 Ring 1" or "GE1/1" detector is designed to fit exactly into the envelope of the older RPC detector.
With respect to the old detector layout, modifications were done to accommodate the new electronics and the new gas system. 
We performed several gas flow simulations\cite{ieee_colafranceschi} focusing on geometrical optimization of the velocity field 
inside the chamber volume. 
We tested different configurations in terms of number of inlets and outlets and spacers geometry. 
For each dataset we estimated the percentage of the chamber where the flow was below a threshold.
The final design was chosen using the feedback from the simulations.
\par
The Triple-GEM detector(in Fig.\ref{ge11cad} black) is fully contained inside an aluminum box and uses large-area 
single-mask GEMs\cite{Villa:2010wj}. 
The production of GEM foils relies on the photolithographic processes, in which the GEM hole pattern is transferred by UV-exposure from flexible
transparent films to a copper-clad polyimide foil laminated with a photoresistive material. 
The foil is chemically etched in order to remove copper from the holes, but not from where the 
photoresist still masks the copper. Finally the polyimide is also etched. 
This process uses two masks for patterning the top and the bottom of the foil, this is an issue 
for the big foil because the alignment, performed manually, between the masks, becomes critical once the foil dimension exceeds $40cm$.
The use of one single mask has been introduced to overcome this problem.
The single mask foils have the same dimensions of the double mask foils; they are made of $50 {\rm {\mu m}}$ thick kapton sheet with $5 {\rm {\mu m}}$ copper clad on both sides. 
The single mask technology is mature and a prototype of the TOTEM T1\cite{Villa:2010wj} detector has been already built for a possible future upgrade. 
\par
Between each GEM foil there is a $2\rm{mm}$ glued spacer (in Fig.\ref{ge11cad} blue); on the right the HV divider is placed (in Fig.\ref{ge11cad} yellow) 
which provides voltage to different sectors of each foil.
The gas mixture enters from the short side via grooves cut (in Fig.\ref{ge11cad} green) in the aluminum frame.
The adopted electronics, the TOTEM VFAT\cite{Aspell:2008zz}, is installed in the openings (in Fig.\ref{ge11cad} violet) that allow connections between the VFAT chips and the readout plane (in Fig.\ref{ge11cad} pink).
  \begin{figure}[H]
  \centering
  \includegraphics[width=3in]{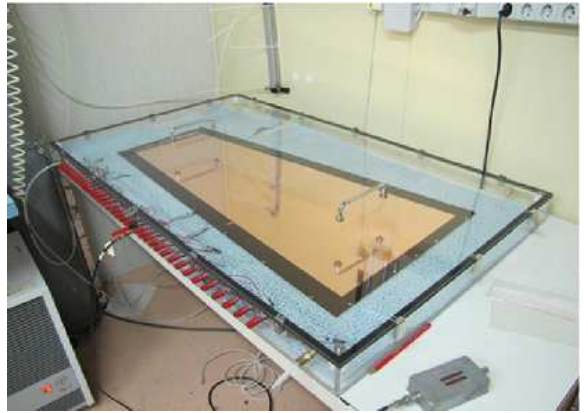}
  \caption{GE1/1 Drift plane under HV test in gas box.}
  \label{drift}
  \end{figure}

The drift electrode (Fig.\ref{drift}), which is part of the chamber envelope itself, 
is produced by gluing a $300 \rm{\mu m}$ kapton layer with $5 \rm{\mu m}$ copper cladding to a $3\rm{mm}$ aluminum plate.
The drift electrode was tested in open air and nitrogen and showed a leakage current less than $20 \rm{nA}$.

\par
 
  \begin{figure}[H]
  \centering
  \includegraphics[width=3in]{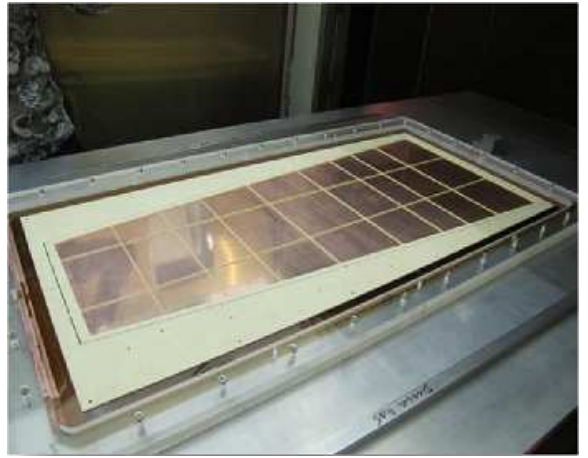}
  \caption{GE1/1 First full-size foil produced.}
  \label{gemfoil}
  \end{figure}
\begin{figure}[H]
  \centering
  \includegraphics[width=3in]{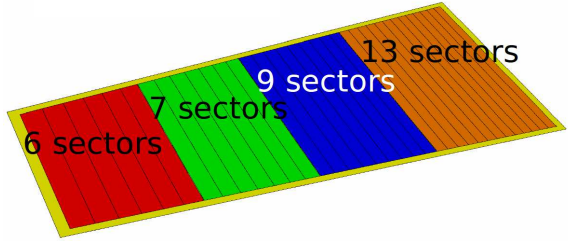}
  \caption{GE1/1 High Voltage sectors.}
  \label{gemsectors}
  \end{figure}

Fig.\ref{gemfoil} shows our first full-size trapezoidal GEM foil with these dimensions: $990\rm{mm} \times (220-455)\rm{mm}$. In Fig.\ref{gemsectors} the sketch of the chosen sectorization of the foil is presented. 
Each sector, out of the 35, covers $100 {\rm cm^2}$ so that the discharge probability (measured at the RD51 lab)
 is about $10^-$$^6$ for a gain of $2 \cdot 10^4$.
Before installing the foils inside the detector, they were subjected to thermal stretching using a special oven. The temperature was maintained at 37${^\circ}C$ for 24 hours. 
Further studies on the stretching process are going on at Florida Tech and Frascati laboratories.
The Florida Tech CMS muon group developed a new, cost-effective technique for thermal stretching\cite{florida} of GEM foils via infrared heating so that the foils can 
be continuously kept under clean room conditions. 
At Frascati, a study of large GEM detectors from the structural point of view has been started\cite{raffone}. 
It shows that for a trapezoidal CMS GEM foil of $1040\rm{mm} \times (345 - 530)\rm{mm}$ the sag due to its own 
weight is about $28.6 \rm{\mu m}$ (electrostatic loads not included) for a tensioning of 1 N/mm; the related stresses are lower than the yield only in a biaxial load. 
The level of accuracy  of these results is very good when compared to some theoretical and experimental results\cite{raffone}.
\par
\begin{figure}[H]
  \centering
  \includegraphics[width=2.5in]{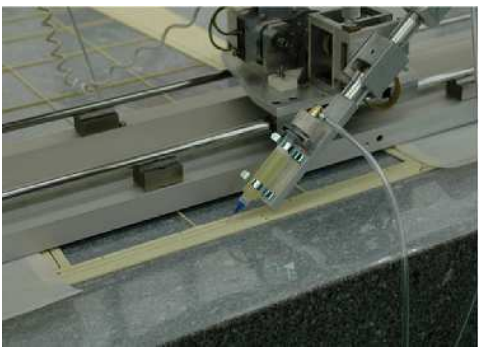}
  \caption{GE1/1 Gluing spacers on the foil with the glue gun.}
  \label{glueing}
  \end{figure}

After the stretching process is completed, the foil is ready to be glued together with the frame (Fig.\ref{glueing}). 
Finally, the foil plus the frame are again placed inside the oven to cure the glue.
While the frame is necessary to keep the foils in place, we performed studies with the purpose of avoiding the stretching process and using any frame at all to have a bigger active area. 
During the test beam campaigns we have tried using a honeycomb structure in between the foils\cite{tytgat_ieee} which is supposed to behave as a frame to maintain the proper distance 
between foils; it was proved 
that this prototype containing this honeycomb worked without problems albeit with a reduced efficiency due to the geometrical influence of the honeycomb walls. Further studies are going on 
to increase the honeycomb cell size to reduce the efficiency loss to a low reasonable value. 
Another strategy to avoid the GEM spacer frames could be to tension the GEM foil  sufficiently by adopting the well tested stretching technique used at Frascati; there this process 
is performed mechanically with accurate control of the applied force\cite{Bencivenni:2009zz}. 
For the GE1/1 prototype, every foil is tested before and after the stretching process; we perform a careful optical inspection and a sector-by-sector HV test
 increasing the voltage up to 500V measuring the current, which we expect to be of the order of few nA. This optical inspection is extremely important because 
simple dust could dramatically affect chamber operations. Because of this, we always operate in a clean room.
\par
\begin{figure}[H]
  \centering
  \includegraphics[width=3in]{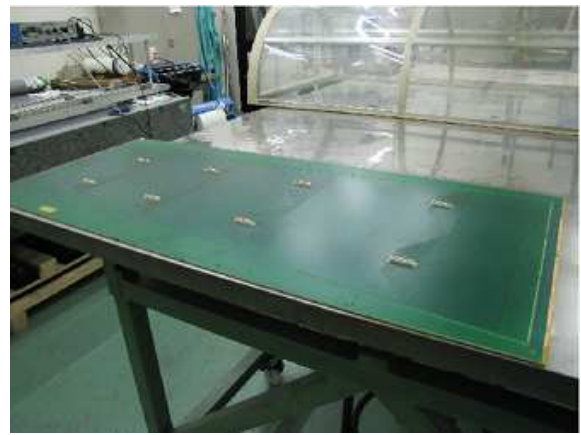}
  \caption{GE1/1 PCB readout plane.}
  \label{readout}
  \end{figure}

The readout PCB (Fig.\ref{readout}) is divided into 4 $\eta$ partitions and each partition has 2 VFAT chips. Each VFAT chip is able to read 128 channels, so for every sector we have 256 channels. 
The strip pitch is varying along the longest chamber dimension from 0.8 mm to 1.6 mm. 
Several noise studies have been done with the same electronics used at the test beam with small size detectors for understanding and debugging the detector. 
The low noise level we achieved made it possible to operate the full-size prototype with approximately the same threshold, and 
therefore operating conditions, used for the small prototype.
\par
\begin{figure}[H]
  \centering
  \includegraphics[width=3.5in]{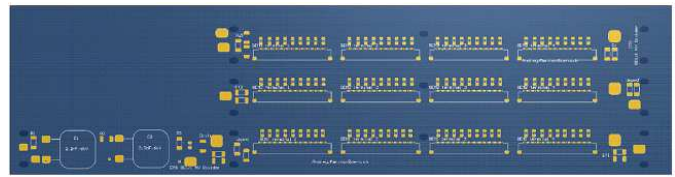}
  \caption{HV divider.}
  \label{divider}
  \end{figure}

Fig.\ref{divider} shows the layout of the HV divider board which provides different voltages to the GEM foils.
It is made using HV SMD resistors and has a built-in RC Filter which cuts the intermediate frequency from the CAEN 1527 power supply, as GEM detectors are sensitive to HV power supply fluctuations.
For the readout electronics we adopted the VFAT(TOTEM) chip because of its capabilities of tracking and fast triggering \cite{Alfonsi:2004jm}, which make it very suitable for our application. 
The VFAT is a digital on/off chip for tracking and triggering with an adjustable threshold for each of the 128 channels; it uses $0.25 \rm{\mu m}$ CMOS technology and its trigger 
function provides programmable “fastOR” information based on the region of the sensor hit. 
For prototype testing we used the front-end 
electronics developed by INFN 
(Siena and Pisa)\cite{Aspell:2008zz}, based on 
the TOTEM VFAT chip.
\section{Performance}
\begin{figure}[H]
  \centering
  \includegraphics[width=2.0in]{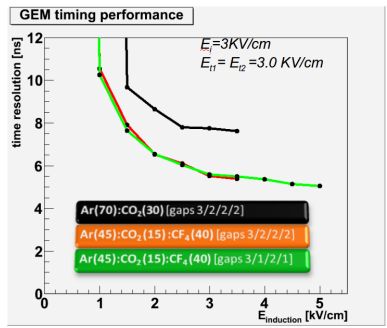}
  \caption{Time resolution of the double mask GEM prototype}
  \label{timeresolution}
  \end{figure}

The performance of our small prototypes has been evaluated during the testbeam that took place at the RD51 setup at H4 beam line (SPS Prevessin) with 150GeV muon/pion beam\cite{tytgat_ieee}. 
The time resolution achieved is around $4.5 \rm{ns}$ as in
Fig.\ref{timeresolution}, with gas mixture $Ar:CO_2:CF_4 (45:15:40)$. 
\par
 \begin{figure}[H]
 \centering
 \subfigure[Efficiency comparison: single and double mask technology]
   {\includegraphics[width=2.0in]{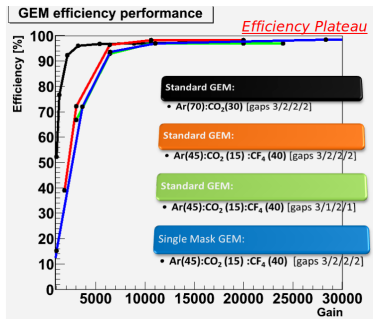}}
 \hspace{5mm}
 \subfigure[Simulated efficiency with LHC bx 25ns]
   {\includegraphics[width=2.0in]{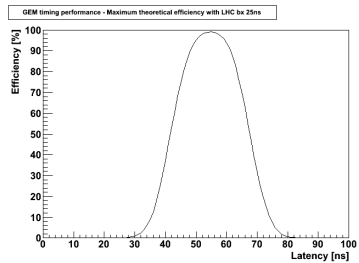}}
 \caption{Efficiency studies}
 \label{efficiency}
 \end{figure}
\begin{figure}[H]
  \centering
  \includegraphics[width=2.5in]{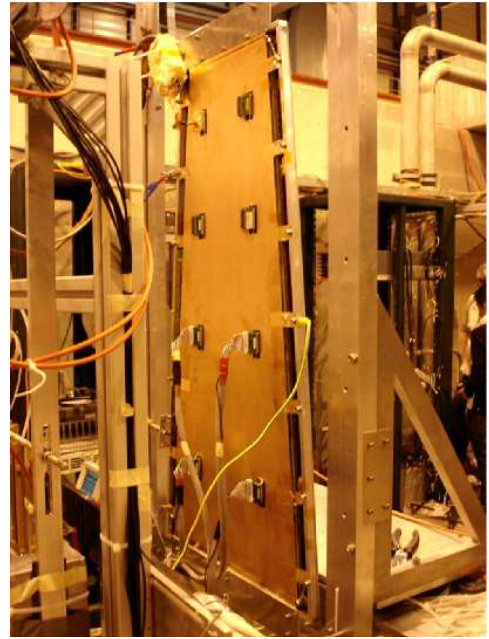}
  \caption{GE1/1 Installed at the Test Beam.}
  \label{ge11tb}
  \end{figure}

In Fig.\ref{efficiency}a we evaluated the efficiency of the prototype built with single mask technology foil and the efficiency of the standard double mask. 
The single mask prototype performs in the same manner as the double mask prototype. This result allows us to replace the standard double mask technology with the single mask, 
which is the only technology that can be suitable for big size GEM foils. 
Using the best time resolution obtained experimentally, we simulated the efficiency for  the LHC bunch-crossing of of $25 \rm{ns}$. 
Fig.\ref{efficiency}b shows the resulting efficiency vs. latency, which is an adjustable parameter of the VFAT chip. 
In October 2010 we completed a test beam using the full-size prototype candidate chamber. 
From an operational point of view, the full-size prototype behaved excellently. 
Fig.\ref{ge11tb} shows the prototype detector mounted on a movable table for a dedicated beam scan along the chamber; the data analysis is currently in progress. 

\section{Summary and Conclusion}

In summary, a full-size working detector with remarkably large dimensions $990\rm{mm} \times (220-455)\rm{mm}$ in the active area has been designed, built, and studied in a beam test. 
Today, the single mask has matured as it profits from a very refined production process which permits careful control of the shapes and dimension of the holes. 
This technology is becoming very suitable even for industrial production, which would decrease cost. 
In 2010 we have completed two beam tests at RD51 setup in the H4 beam line (SPS Prevessin), with small size prototypes; 
we demonstrated that the small prototypes are addressing the requested requirements in terms of high efficiency and gain, stable safe and reliable operation at LHC. 
From the electronic point of view, the detector plus the readout proved to work without any problem under simulated CMS LHC conditions. 
The data analysis is still ongoing and preliminary results will be available soon. Meanwhile, the production of a second full-size prototype is being planned.

\section*{Acknowledgment}
We particularly thank G. Bencivenni (Frascati) for discussions and advice.


%

\end{document}